\documentclass[a4paper,12pt]{article}

\usepackage[font=large,labelfont=bf]{caption}

\usepackage[final]{pdfpages}

\usepackage{setspace}

\usepackage{lscape}

\usepackage{epsf,amsfonts,amsmath}

\usepackage{epstopdf}

\usepackage[T1]{fontenc} 

\usepackage{booktabs,caption,fixltx2e}

\usepackage{graphicx} 

\usepackage{ifpdf}

 \usepackage{subcaption}

\usepackage{setspace}

\usepackage{ctable}

\usepackage[flushleft]{threeparttable}

\usepackage{caption}

\usepackage{amsmath}

\usepackage{amsfonts} 

\usepackage{xkeyval}

\usepackage[round]{natbib}

\usepackage{longtable} 

\usepackage{pdflscape}

\usepackage[english]{babel} 

\usepackage{inputenc}

\usepackage{multirow}

\usepackage{rotating}

\usepackage{xtab}

\usepackage{sidecap}

\usepackage{float}

\usepackage{natbib}

\usepackage{apalike}

\usepackage{multicol}
\newcommand{\commentout}[1]{}

\newtheorem{result}{Result}

\begin{document}

\title{Gender differences in altruism: \\ Expectations, actual behaviour and accuracy of beliefs} 
\author{Pablo Bra\~{n}as-Garza\thanks{Economics Department, Middlesex University London, Business School, The Burroughs, London NW4 4BT, United Kingdom.}, Valerio Capraro\thanks{Center for Mathematics and Computer Science, Science Park 123, 1098 XG, Amsterdam, The Netherlands.}, Ericka Rasc\'{o}n-Ram\'{i}rez$^*$}
\maketitle

\normalsize

\begin{abstract}
Previous research shows that women are more altruist than men in dictator game experiments. Yet, little is known whether women are expected to be more altruist than men. Here we elicit third-parties' beliefs about dictators' donations conditional on knowing the gender of the dictator. Our data provide evidence of three main findings: (i) women are expected to be more altruist than men; (ii) both men and women have correct beliefs about the level of altruism among men; and (iii) both men and women overestimate the level of altruism among women. In doing so, our results uncover a perception gap according to which, although women are more altruist than men, they are expected to be even more altruist than they actually are.

\end{abstract}

\section{Introduction} \label{sec:Intro}

\begin{multicols}{2}

The exploration of gender differences in decision making has a long tradition in behavioural economics and other social sciences, and has touched several research areas, including risk-aversion, competitive behaviour, and social preferences. 

For example, a classical study by Eckel and Grossman (\citeyear{eckel2002sex}) has shown that women are more risk averse than men, while an equally classical study by Niederle and Vesterlund (\citeyear{niederle2007women}) has shown that women are less competitive than men. In terms of social preferences, results are more mixed: while previous research has not uncovered any obvious gender difference in cooperative behaviour (see \cite{croson2009gender} for a review), experimental studies have repeatedly found that women are, on average, more altruist than men (\cite{bolton1995experimental,eckel1998women,andreoni2001fair,dufwenberg2006gender,houser2009fairness,dreber2014gender,capraro2014good,capraro2015emergence}). More recently, \cite{rand2016social} have extended this line of research by showing, through a meta-analysis of 22 studies, that promoting intuition versus reflection increases altruistic behaviour among women, but not among men, suggesting that women but not men have internalised altruism as their spontaneous reaction. 

Other studies have shown that women tend to be more altruistic than men when investing in human capital for children. For example, women allocate more resources for women's and children's clothing relative to men's clothing (\cite{lundberg1997husbands}), invest more in health and nutrition for children (\cite{duflo2000child}), and spend more on child goods and small scale livestock (\cite{rubalcava2009investments}) than men.  

Despite the vast research in gender differences, the literature has mostly neglected the inverse question of whether people have gender stereotypes in specific decisional settings. Understanding whether people have correct beliefs about others' behaviour is an important question \emph{per se}, because the standard equilibrium analysis assumes that people strategise on their beliefs about their counterparts' behaviour (\cite{camerer2004behavioural}); and it becomes even more important when it comes to gender differences, since one of the dominant explanations for gender differences in decision-making relies on the assumption that the behaviour of men and women is governed by stereotypes regarding their social roles (\cite{eagly1987sex,brescoll2011who}). In sum, understanding whether there is a correspondence between stereotypes of men and women and their actual behaviour is an important question, with potential consequences in economic and psychological modelling.

Given the aforementioned literature showing that women are more altruistic than men, here we ask whether this gender difference in behaviour corresponds to a gender difference in stereotypes. We move a first step into this research area by starting from a simple question: \emph{are women expected to be more altruistic than men?}

To the best of our knowledge, only a handful of papers have approached this question, and mostly did so from a psychological perspective. For example, Heilman and Chen (\citeyear{heilman2005same}) showed that work-related altruism is less optional for women than for men, and Heilman and Okimoto (\citeyear{heilman2007why}) showed that penalties for women's success in male domains result from the perceived violation of gender-stereotypic prescriptions. From an economic perspective, we are aware of only one study devoted to eliciting participants' beliefs about the level of altruism in men and women (\cite{aguiar2009women}). In this lab experiment, subjects were presented with two boxes, A and B, where A contained donations left by men and B contained donations left by women. Subjects were informed that they could choose only one of the two boxes and one donation would be taken at random from the selected box and used to pay them. Results showed that subjects were more likely to select donations from the "women" box, indicating that women were indeed expected to be more generous than men.   

Although it represents an important first step towards understanding whether women are expected to be more altruistic than men, the work by Aguiar et al. (\citeyear{aguiar2009women}) has two important limitations. First of all, while it shows that women are expected to be more generous than men, it does not show whether people have correct beliefs about the behaviour of men and women. Thus, it remains unclear whether people have correct stereotypes regarding each gender's level of altruism. Second, it is \emph{only one study}: the recent outbreak of the replicability crisis (Open Science Collaboration, \citeyear{open2015}) calls for more studies. 

In the current work, we wish to fill these gaps by: (i) replicating the result that women are expected to be more generous than men; and (ii) giving a quantitative version of this result. This allows to answer the question: do men and women fulfil people's expectations about altruistic behaviour?

The rest of the paper is organised as follows. Next section is devoted to methods; section \ref{sec:results} focuses on results and discussion; last section concludes.


\section{Method} \label{sec:method}

\subsection{Subject pool}

Subjects were living in the US at the time of the experiment and were recruited using Amazon Mechanical Turk (\cite{paolacci2010running,horton2011online,mason2012conducting,paolacci2014inside}) to play a standard Dictator Game (\cite{kahneman1986fairness,forsythe1994fairness}). 

In the Dictator Game, one player acts in the role of the dictator and the other one in the role of the receiver. Dictators are given a certain amount of money and are asked how much, if any, they want to give to the receiver. Receivers have no choice and only get what the dictators decide to give. Since dictators have no incentives to give money, a payoff-maximising dictator would donate nothing. For this reason, dictators' donations are taken as a measure of individual's altruism, or inequity aversion (\cite{fehr1999theory,bolton2000erc,branas2006poverty,branas2007promoting,charness2008s}). 

\subsection{Protocol }

In our experiment, subjects were randomly divided between dictators and receivers. \\
\\
\textbf{Dictators:} They were given $\$0.20$ and were asked to decide how much, if any, to give to the receiver. Before making their decision, dictators were asked two comprehension questions. Specifically, they were asked which choice would maximise their payoff and which choice would maximise the receiver's payoff. Subjects failing any comprehension question were automatically excluded from the survey. This screening procedure had the effect that we had fewer dictators ($N=456$) than receivers ($N=530$). Thus, the computation of receivers' payoffs is not straightforward, since there is no one-to-one correspondence between dictators and receivers. To address this problem, receivers were sequentially paired with a randomly selected dictator; in case a dictator was already used to pay another receiver, we paid the current recipient `out of our pocket', and not using the donation of that dictator, because that donation had already been used. This procedure is doable on Amazon Mechanical Turk, because participants are matched only after the end of the experiment.\\
\\
	\textbf{Receivers:} A part from potentially receiving money from dictators, receivers played also as guessers. Specifically, they were asked to predict the donation that another dictator would make to another receiver. They would receive, on top of the actual donation, $\$0.20$ reward for correct guesses. Since they do not guess their own donation there is no opportunity to hedge (\cite{branasguess}). To elicit recipients' expectations, we designed four treatments: 

\begin{itemize}
\item[O$_{\text{n}}$:]  recipients were presented with the same screenshots shown to dictators and they were asked to guess the dictator's decision ($N = 134$);
\item[O$_{\text{mow}}$:] was identical to  O$_{\text{n}}$ with the only difference that recipients were informed that the dictator was either a man or a woman ($N=140$). 
\item[O$_{\text{m}}$:] was identical to O$_{\text{n}}$ with the only difference that recipients were informed that the dictator was a man ($N=124$);
\item[O$_{\text{w}}$:] was identical to O$_{\text{m}}$ with the only difference that recipients were informed that the dictator was a woman ($N=132$).
\end{itemize}

We need both  O$_{\text{n}}$ and O$_{\text{mow}}$ baselines for two reasons: on the one hand, by comparing O$_{\text{m}}$ and O$_{\text{w}}$ with O$_{\text{mow}}$, separately, we may investigate the effect of making one particular gender salient versus making both genders salient; on the other hand, by comparing dictators' donations with O$_{\text{n}}$, we can explore whether people have correct beliefs about the level of altruism in anonymous strangers.




\section{Results}\label{sec:results}


\subsection{Descriptive statistics}
A total of 986 subjects (56\% men, mean age = 34.5 years) participated in our experiment. The average donation was 27.3\% of the total endowment, which is very close to the average donation reported in Engel's meta-analysis of 616 Dictator game experiments conducted in the standard physical laboratory (28.3\%, \cite{engel2011dictator}). This confirms the reliability of data collected on Amazon Mechanical Turk using very small stakes, a fact that was already observed in the context of the Dictator Game by d'Adda et al. (\citeyear{dadda2015push}). Although the pie size was $\$0.20$ data are normalised such that the donations correspond to 0-10. Next we pass to the analysis of treatment effects.

\subsection{Gender framed vs non-framed treatments}

As a preliminary step we start by looking at framing effects on recipients' beliefs. Both treatments O$_{\text{n}}$ and O$_{\text{mow}}$ report similar averages ($2.79$ and $3.16$, resp.). Table \ref{tab:guesses} shows no significant differences between O$_{\text{n}}$ and O$_{\text{mow}}$ (t-test, $p = 0.24$; z-test, $p=0.23$). Similarly, we do not find significant differences between O$_{\text{m}}$ $\cup$ O$_{\text{w}}$ and O$_{\text{mow}}$ (t-test, $p = 0.89$; z-test, $p=0.84$). Hence, the sum of `men' and `women' frames is equal to the treatment in which `both' genders are mentioned.

Therefore, we may conclude that \emph{mentioning `gender' does not frame recipients expectations}. 

\subsection{Are women expected to be more generous than men?}

To answer this question we compare treatment O$_{\text{m}}$ with O$_{\text{mow}}$ and treatment O$_{\text{w}}$ with O$_{\text{mow}}$. Figure \ref{fig:exp} shows the distribution of beliefs by treatment. Figures 1\emph{a}, 1\emph{b} and 1\emph{c} show the histograms for O$_{\text{m}}$, O$_{\text{mow}}$ and O$_{\text{w}}$, respectively. While the modal values for expected behavior of males is 0 (giving nothing) the modal for women (i.e., O$_{\text{w}}$) and, to a lesser extent, for women or men (i.e., O$_{\text{mow}}$) is the equal split. Average values reflect the same result: the mean expected altruism in O$_{\text{m}}$  is 2.33, while the mean for O$_{\text{mow}}$ is 3.18 (t-test, $p = 0.01$; z-test, $p = 0.01$). Conversely, when the dictator is a  `woman', the mean expected generosity in O$_{\text{w}}$ is 4.05, which is significantly larger than the mean for O$_{\text{mow}}$ (t-test, $p = 0.01$; z-test, $p = 0.00$). Comparing the expected level of generosity among males and females, O$_{\text{w}}$ vs O$_{\text{m}}$, we observe than the average and median differences are 1.72 and 4 units, respectively. The top part of Table \ref{tab:guesses} shows the relevant tests.

Figure $1a$ to Figure $1c$ provide visual evidence that we can reject the hypothesis that men are expected to be as altruistic as the average person and that women are expected to be as altruistic as the average person. 

Figure $1d$ focuses on the CDFs (cumulative distribution functions). While males' CDF is closer to the top right -- more selfish -- Females CDF is closer to the bottom left --more generous. It is easy to check that O$_{\text{w}}$ stocastically dominates O$_{\text{m}}$ which is consistent with the test shown in Table \ref{tab:guesses}. The entire distribution of O$_{\text{w}}$ is always toward the right of the rest of distributions. In sum, women are expected to me more altruistic than men.

\begin{result}
Women are expected to be more altruistic than men.
\end{result}



\commentout{
\begin{figure}[H]
\caption{Expected behavior for men, women and both} \label{fig:exp}

\begin{subfigure}{0.5\textwidth}
\includegraphics[width=\linewidth]{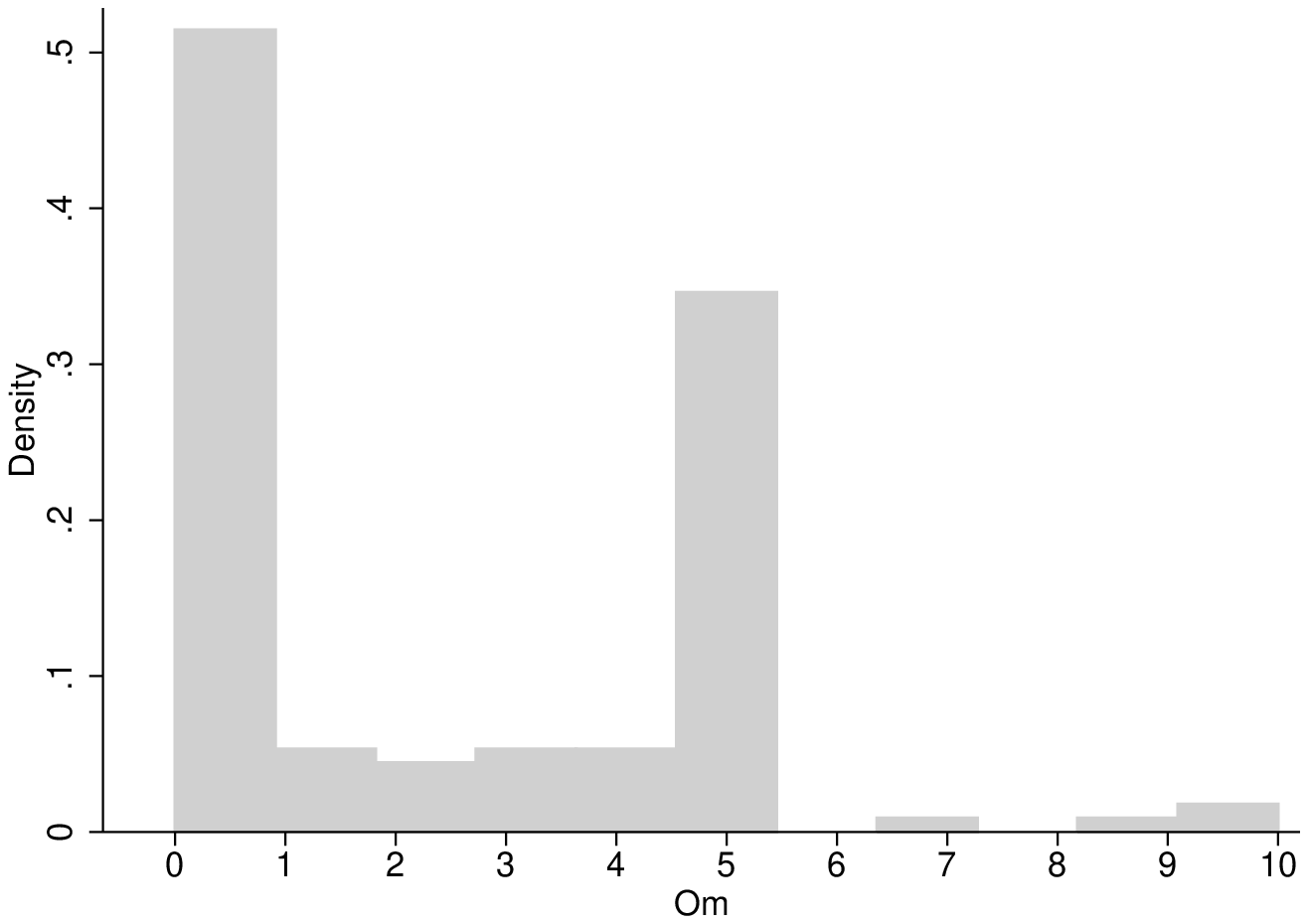}
\caption{Expectations for men [O$_{\text{m}}$]} \label{fig:1a}
\end{subfigure}
\hspace*{\fill} 
\begin{subfigure}{0.5\textwidth}
\includegraphics[width=\linewidth]{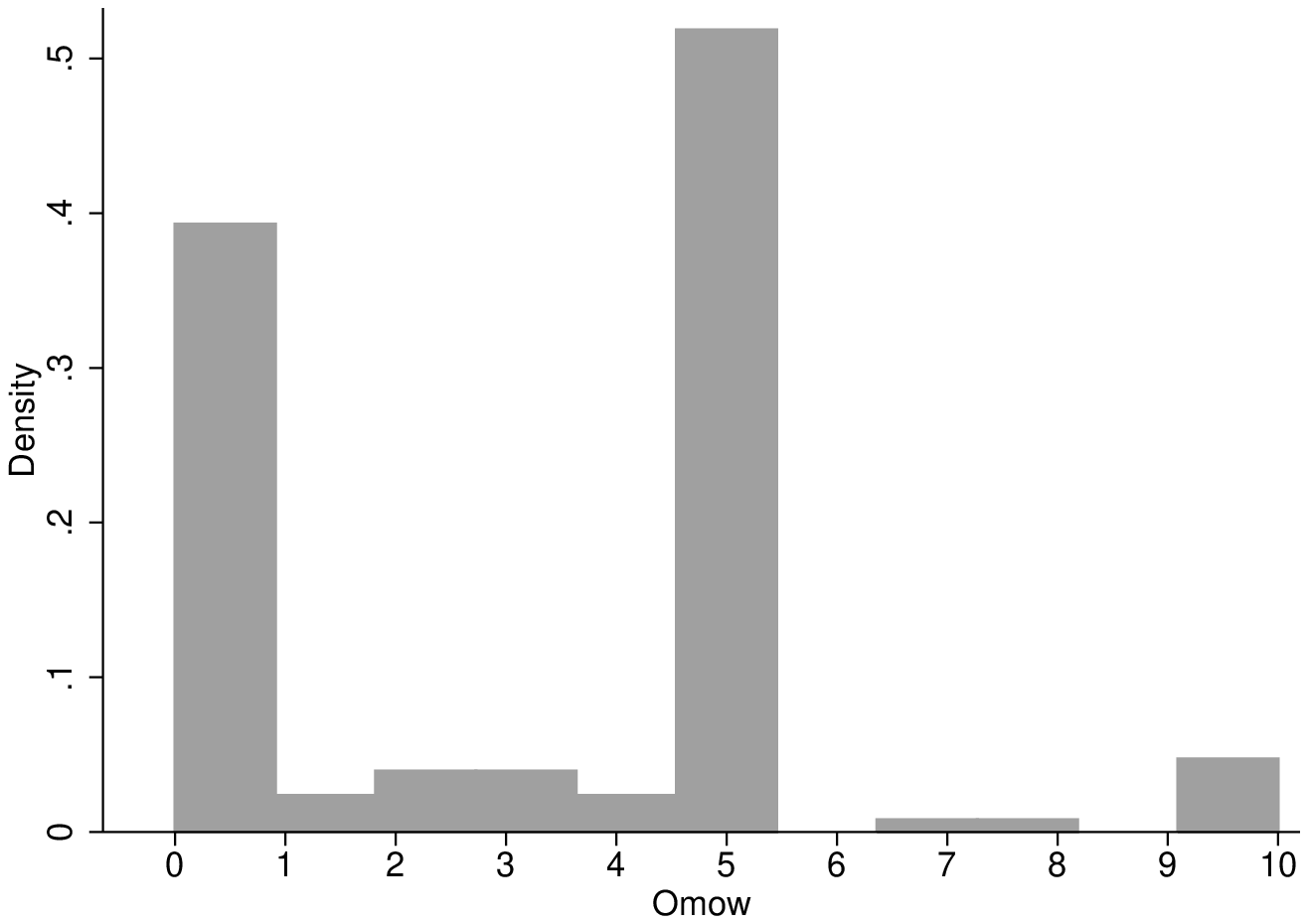}
\caption{Expectations for either men or women [O$_{\text{mow}}$]} \label{fig:1b}
\end{subfigure}\\
\begin{subfigure}{0.5\textwidth}
\includegraphics[width=\linewidth]{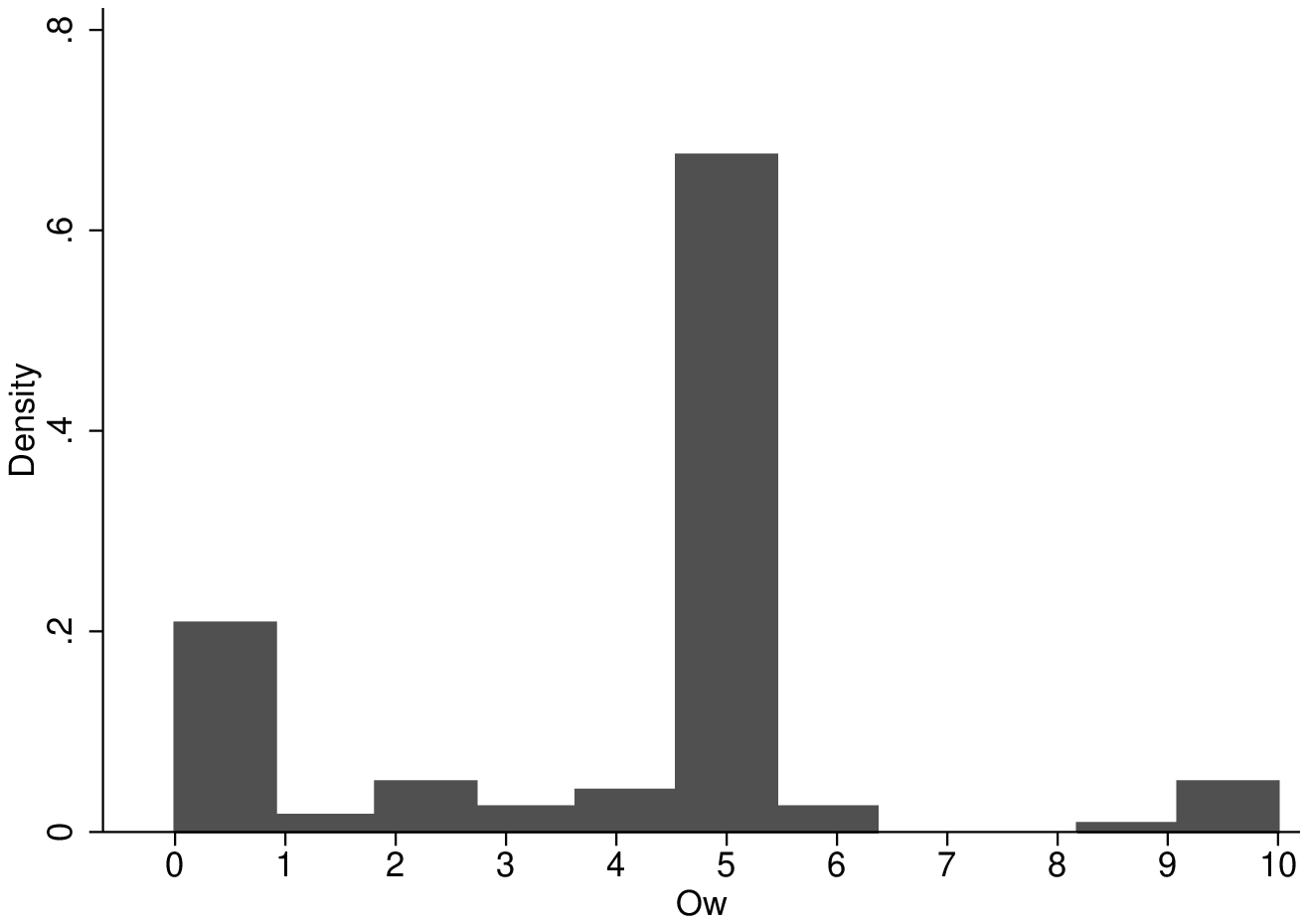}
\caption{Expectations for women [O$_{\text{w}}$]} \label{fig:1c}
\end{subfigure}
\hspace*{\fill} 
\begin{subfigure}{0.5\textwidth}
\includegraphics[width=\linewidth]{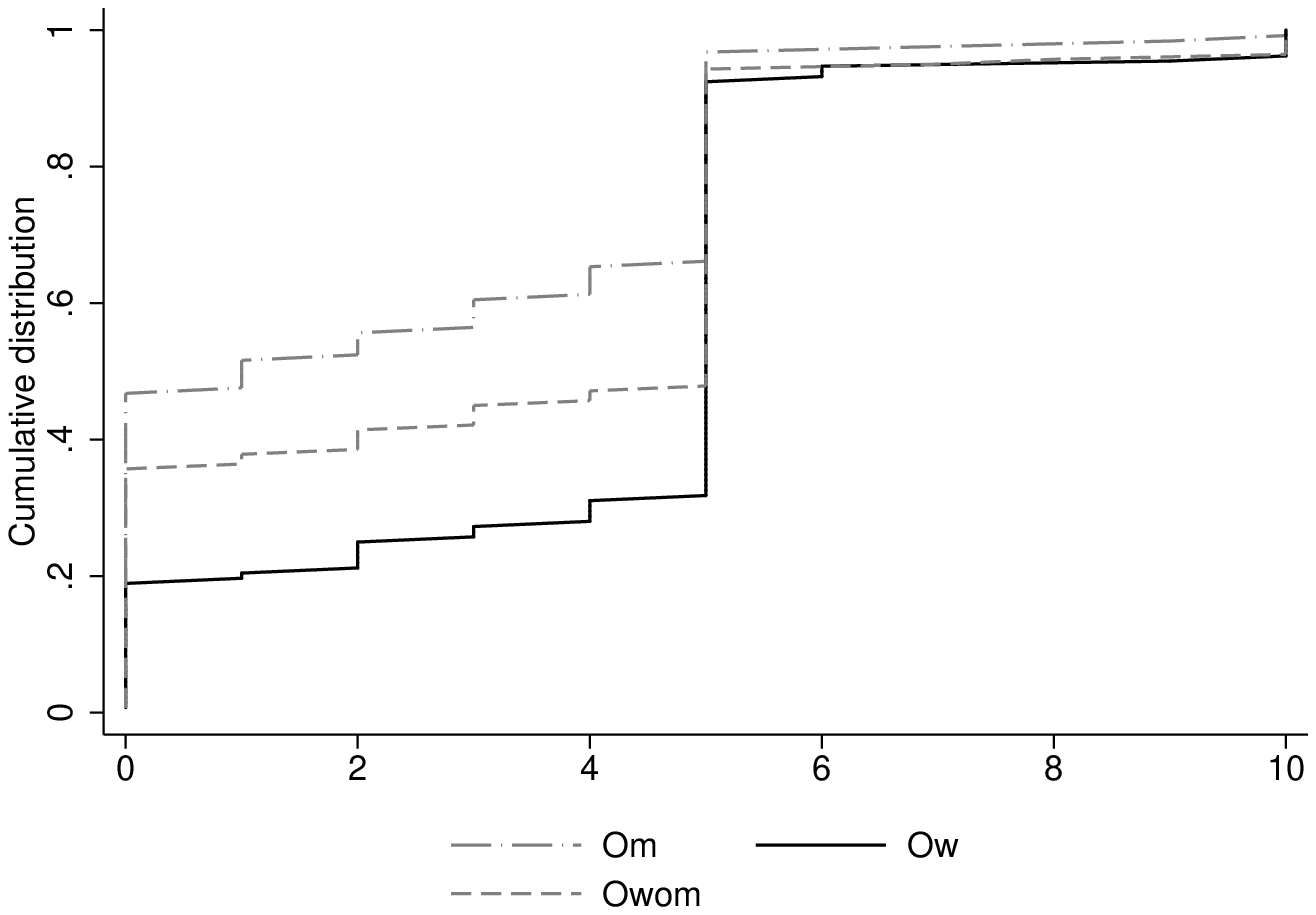}
\caption{CDFs for O$_{\text{m}}$, O$_{\text{mow}}$ and O$_{\text{w}}$} \label{T2vsT3.eps}
\end{subfigure}
\end{figure}
}


\subsection{Are women actually more generous than men?}

In the previous subsection, we have shown that women are expected to be more altruistic than men in a Dictator Game. Is this expectation grounded or not? 

Figure 2\emph{a} and Figure 2\emph{b} respectively compare the distribution of donations for both men and women, and provides visual evidence that women are, on average, more altruist than men (means: $3.04$ vs $2.49$; t-test, $p = 0.03$; z-test, $p = 0.01$). In fact, giving nothing is the modal value for males (49.6\% gave 0) while giving the equal split is the modal value for women  (48.3\% gave half). 

In sum, not only women are expected to be more generous than men, but they are \emph{de facto} more generous than men.


\begin{result}
Women are more altruistic than men.
\end{result}



\commentout{
\begin{figure}[H]
\caption{Actual behaviour: men vs women} \label{fig:2}

\begin{subfigure}{0.5\textwidth}
\includegraphics[width=\linewidth]{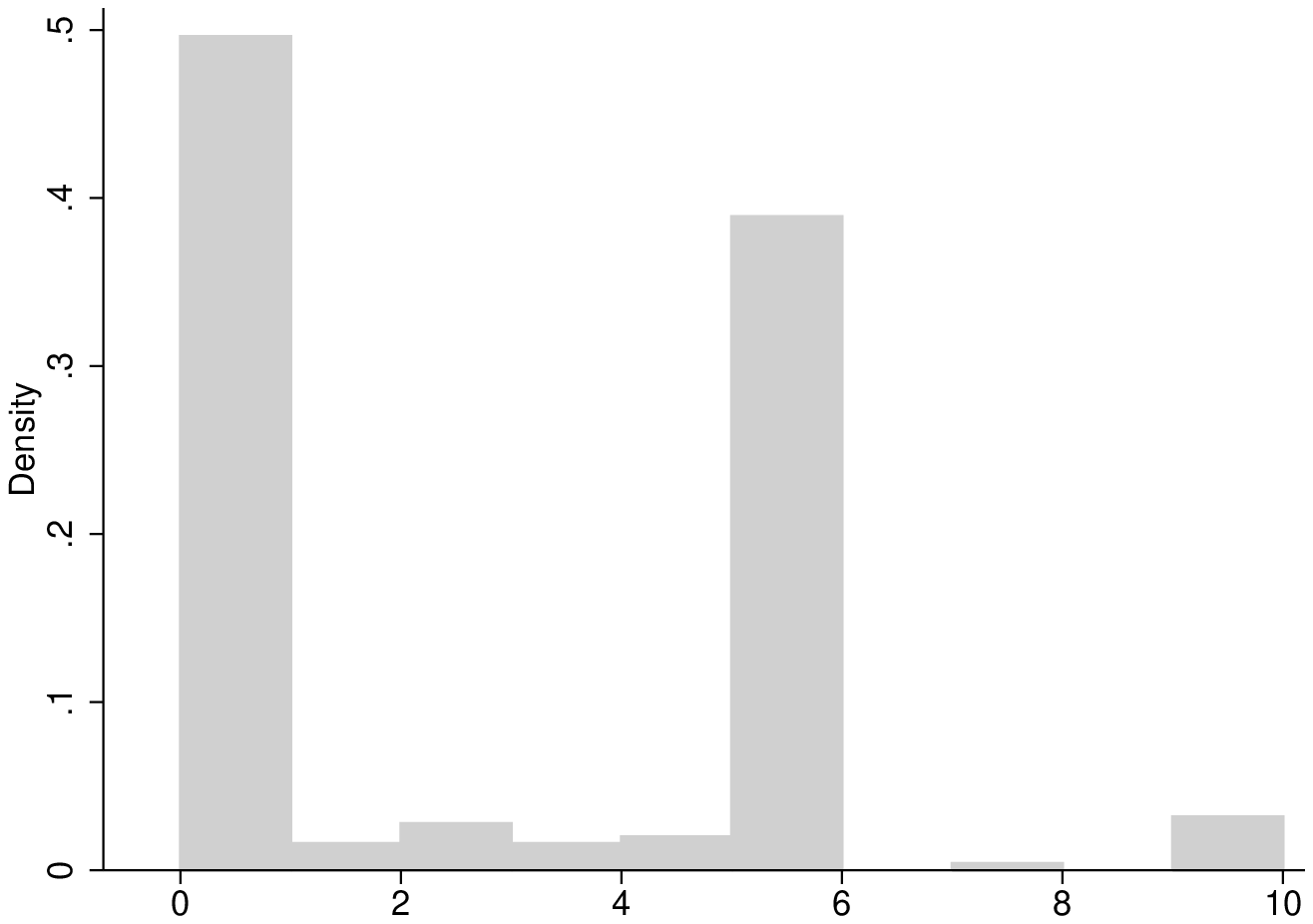}
\caption{Actual behaviour of men [Dm]} \label{fig:2a}
\end{subfigure}
\hspace*{\fill} 
\begin{subfigure}{0.5\textwidth}
\includegraphics[width=\linewidth]{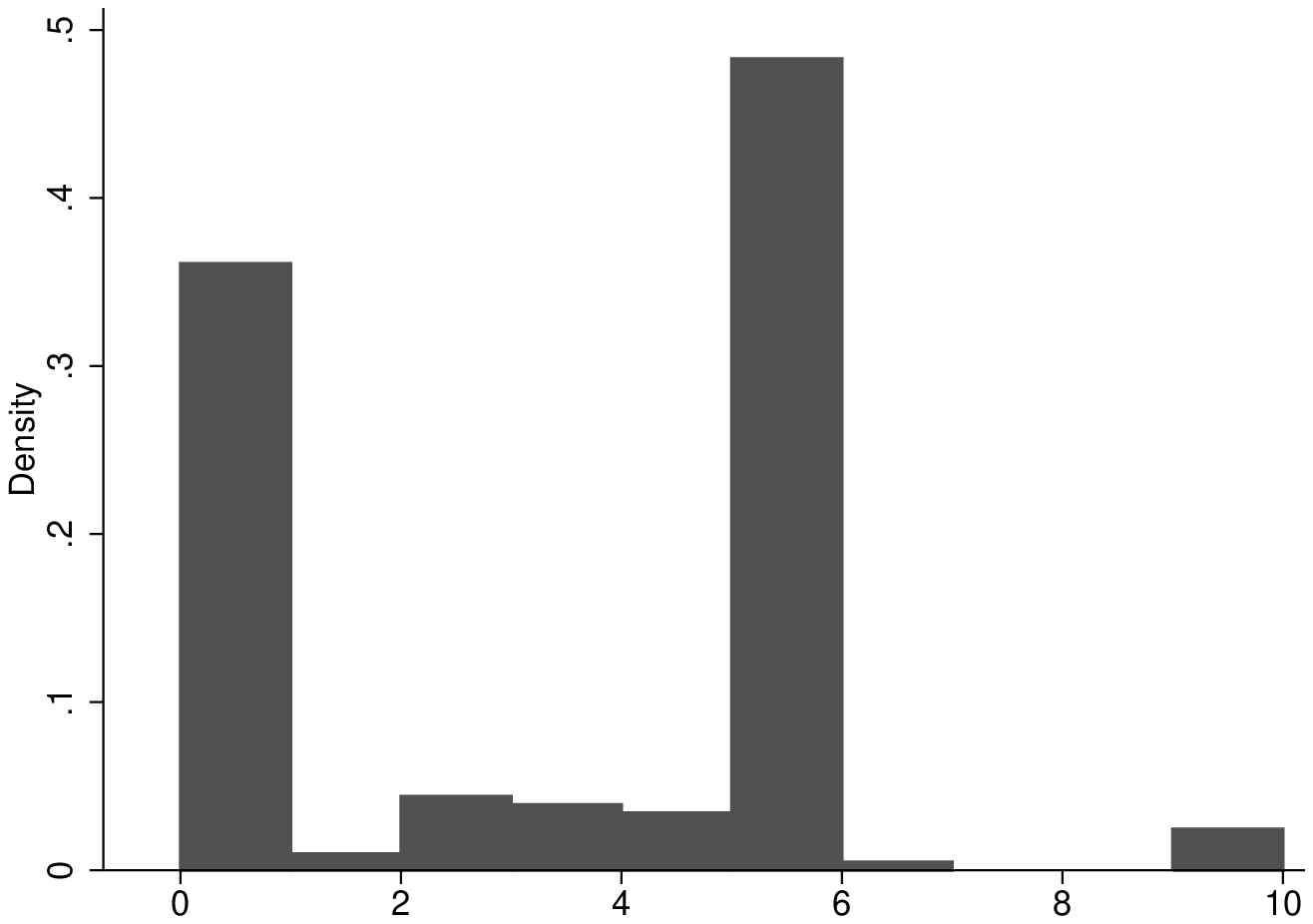}
\caption{Actual behaviour of women [Dw]} \label{fig:2b}
\end{subfigure}\\

\end{figure}
}


\subsection{Do subjects have correct beliefs about each gender's average level of generosity?}
	
In the previous subsections, we have shown that women are expected to be more altruistic than men and that this expectation is grounded, in the sense that women are actually more altruistic than men. Now, we ask whether people have correct beliefs about each gender's average level of generosity.


We begin by observing that subjects have, on average, correct beliefs about the average level of altruism. Specifically, the mean level of altruism across the experiment (both males and females) is 2.735, while the mean level of expected generosity in the O$_{\text{n}}$ condition is 2.798 (t-test, $p = 0.81$; z-test, $p = 0.83)$, see Table 1 bottom). Hence \emph{subjects have correct beliefs about average level of generosity}, which in turn means that we do not observe either wishful thinking or pessimism. 



\commentout{
\begin{figure}[H]
\caption{Accuracy of beliefs: Expected vs actual behaviour} \label{fig:2}

\begin{subfigure}{0.5\textwidth}
\includegraphics[width=\linewidth]{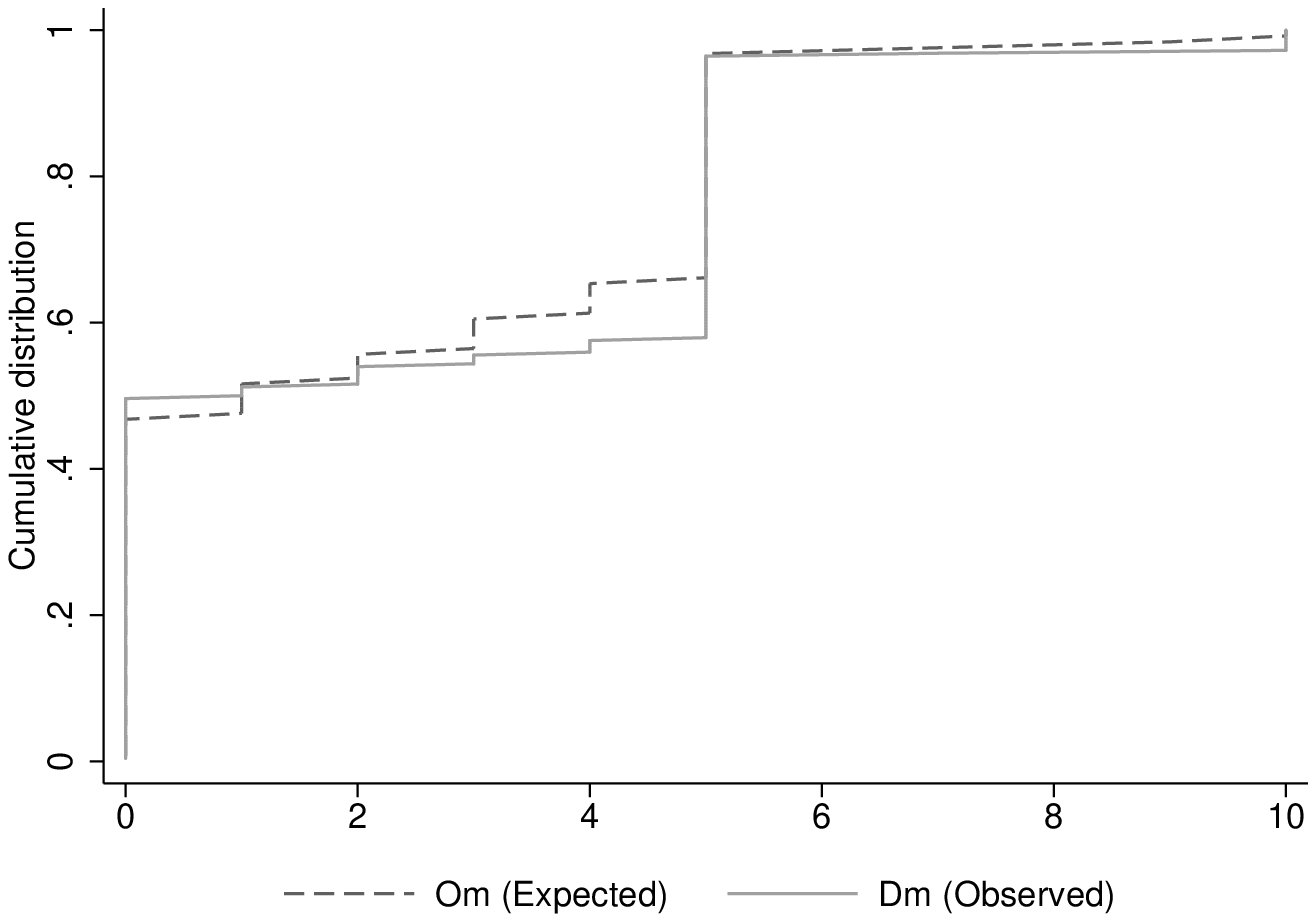}
\caption{Accuracy of beliefs for men} \label{fig:3a}
\end{subfigure}
\hspace*{\fill} 
\begin{subfigure}{0.5\textwidth}
\includegraphics[width=\linewidth]{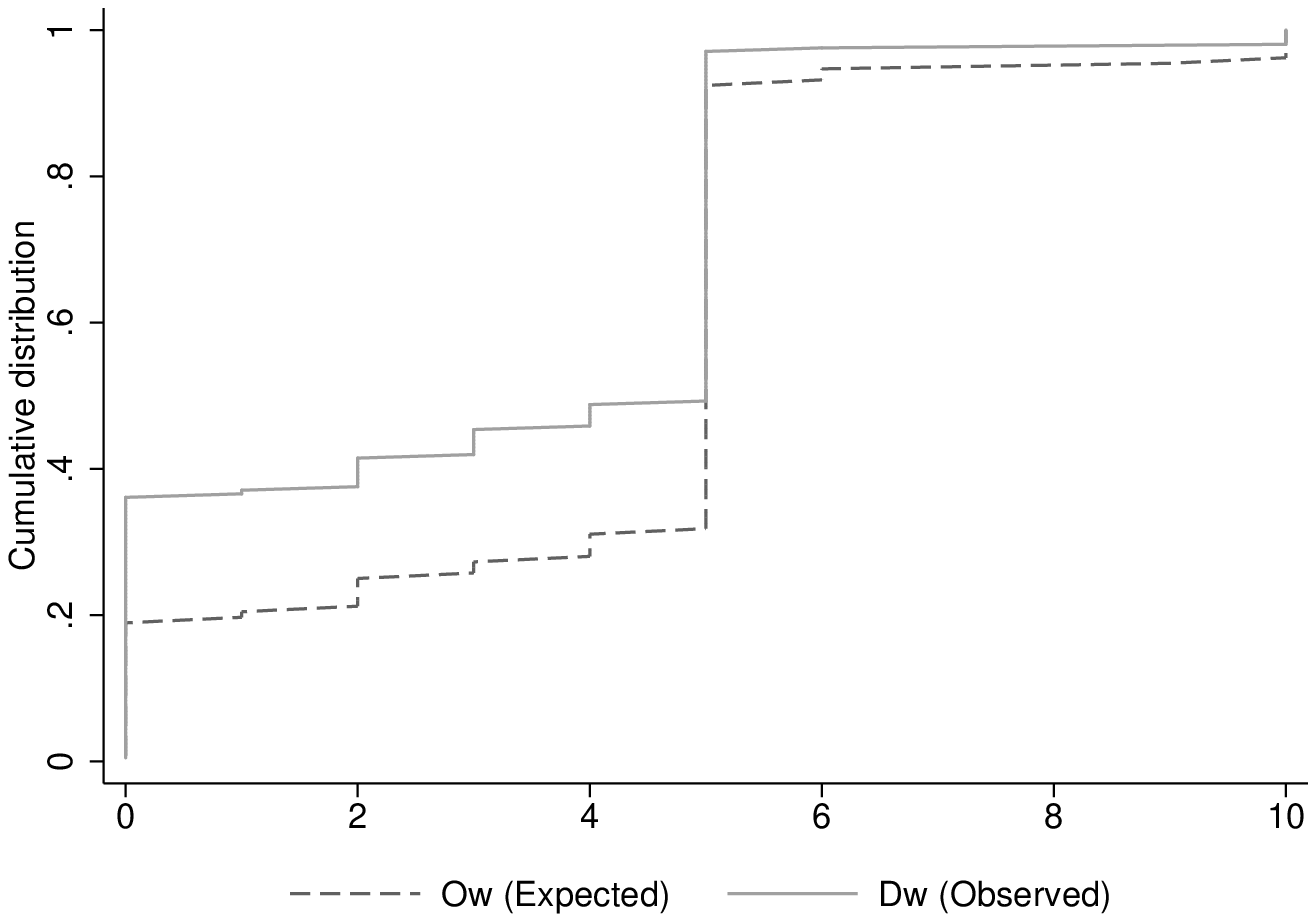}
\caption{Accuracy of beliefs for women} \label{T2vsT3.eps}
\end{subfigure}
\end{figure}
}


Next we analyse whether subjects have correct beliefs about men's average level of altruism. Figures $3a$ analyses accuracy of beliefs for men and shows that there is no discrepancies since both expectations and actual behavior are almost identical (CDFs are on parallel).

Controlling for the gender of the recipient, we also find that both men and women have, on average, correct beliefs about men's level of altruism (t-test, guess by men $ p = 0.21 $ and guess by women $ p = 0.60$; z-test, guess by men $ p = 0.22 $ and guess by women $ p = 0.44$, see Table 1 bottom). 

\begin{result}
Both men and women have correct beliefs about average level of generosity in men.
\end{result}

However Figure $3b$ shows strong discrepancies between current behaviour and expectations for women: females are not as generous as they are expected to be (O$_{\text{w}}$ CDF dominates the  D$_{\text{w}}$ CDF). 

This remains true also after controlling for gender. Both men and women overestimate women's average level of generosity (t-test, both p-values $<0.03$; z-test, both p-values $<0.02$, see Table 1).

\begin{result}
Both men and women overestimate the level of generosity in women.
\end{result}

\end{multicols}

\begin{table}[ht!] 
\centering \textbf{\caption{Hypothesis testing}\label{tab:guesses}}~\\ 
\footnotesize
	\begin{threeparttable}
\begin{tabular}{@{}l*{8}{c}@{}} 
\toprule
\textbf{Hypothesis}	&	\multicolumn{3}{c}{\textbf{Parametric Tests}}	&	&	\multicolumn{3}{c}{\textbf{Non-Parametric Tests}}	\\

\cmidrule{2-4} \cmidrule{6-8} 

										&	\textbf{Difference}	&	\textbf{T-test}	&	\textbf{P-Value}	&	&	\textbf{Difference}	&	\textbf{Z-tests}	&	\textbf{P-value}	\\
										&	\textbf{in Means}	&										&										&	&	\textbf{in Medians}	&	&		\\

\midrule				
O$_{\text{n}}$= O$_{\text{mow}}$	&	-0.38	&	-1.17	&	0.24	&	&	-2	&	-1.19	&	0.23	\\
O$_{\text{m}}$ $\cup$ O$_{\text{w}}$ = O$_{\text{mow}}$	&	0.04	&	0.14	&	0.89	&	&	0	&	0.20	&	0.84	\\
O$_{\text{m}}$ = O$_{\text{mow}}$	&	-0.85	&	-2.58	&	0.01	&	&	-4	&	-2.59	&	0.01	\\
O$_{\text{w}}$ = O$_{\text{mow}}$	&	0.87	&	2.77	&	0.01	&	&	0	&	2.89	&	0.00	\\
O$_{\text{m}}$ = O$_{\text{w}}$	&	1.72	&	5.50	&	0.00	&	&	4	&	5.51	&	0.00	\\
\midrule														
D= O$_{\text{n}}$	&	-0.06	&	-0.25	&	0.81	&	&	0	&	-0.22	&	0.83	\\
D$_{\text{m}}$  = O$_{\text{m}}^{\text{m}}$	&	0.44	&	1.26	&	0.21	&	&	1	&	1.24	&	0.22	\\
D$_{\text{m}}$  = O$_{\text{m}}^{\text{w}}$	&	-0.20	&	-0.53	&	0.60	&	&	-1	&	-0.78	&	0.44	\\
D$_{\text{w}}$  = O$_{\text{w}}^{\text{m}}$	&	-0.81	&	-2.25	&	0.03	&	&	0	&	-2.28	&	0.02	\\
D$_{\text{w}}$  = O$_{\text{w}}^{\text{w}}$	&	-1.27	&	-3.92	&	0.00	&	&	0	&	-3.57	&	0.00	\\
D$_{\text{w}}$  = D$_{\text{m}}$ 	&	0.55	&	2.21	&	0.03	&	&	4	&	2.36	&	0.02	\\	

\bottomrule				
\end{tabular}
\begin{tablenotes}
	\small
 	\item \footnotesize Note: t-tests assume unequal variances per treatment and normality of the distribution of differences in means; z-tests correspond to Mann-Whitney-Wilcoxon non-parametric tests.	D$_{\text{m}}$ (D$_{\text{w}}$) refers to men (women) dictators; D to $any$ dictator.
\end{tablenotes}
	\end{threeparttable}
	\end{table}

\section{Conclusion} \label{sec:concl}

\begin{multicols}{2}
Here we have used Dictator Game experiments to measure people's expectations about dictators' level of generosity, conditional on knowing the gender of the dictator. Our data provide evidence of three major results: (i) women are expected to be more generous than men (replicating \cite{aguiar2009women} results); (ii) both men and women have correct beliefs about the mean level of generosity among men; (iii) both men and women overestimate the level of generosity among women.

In doing so, our experiment uncovers a perception gap according to which, although women are more altruist than men, they are expected to be even more altruistic than they actually are. This result is particularly puzzling since it regards also women: while women have correct beliefs about the level of altruism in men, they overestimate the level of altruism in other women.

We hope that future research can shed light on the ultimate origin of this perception gap and on the potential psychological and economic consequences that can have on women's and men's behaviour.

\end{multicols}

\pagebreak	
	
\section*{}

\bibliographystyle{apalike}

\bibliography{biblio}

\begin{thebibliography}{}

\bibitem[Aguiar et~al., 2009]{aguiar2009women}
Aguiar, F., Bra{\~n}as-Garza, P., Cobo-Reyes, R., Jimenez, N., and Miller,
  L.~M. (2009).
\newblock Are women expected to be more generous?
\newblock {\em Experimental Economics}, 12(1):93--98.

\bibitem[Andreoni and Vesterlund, 2001]{andreoni2001fair}
Andreoni, J. and Vesterlund, L. (2001).
\newblock Which is the fair sex? {G}ender differences in altruism.
\newblock {\em The Quarterly Journal of Economics}, 116:293--312.

\bibitem[Bolton and Katok, 1995]{bolton1995experimental}
Bolton, G.~E. and Katok, E. (1995).
\newblock An experimental test for gender differences in beneficent behavior.
\newblock {\em Economics Letters}, 48(3):287--292.

\bibitem[Bolton and Ockenfels, 2000]{bolton2000erc}
Bolton, G.~E. and Ockenfels, A. (2000).
\newblock {E}{R}{C}: {A} theory of equity, reciprocity, and competition.
\newblock {\em The American Economic Review}, 90:166--193.

\bibitem[Bra{\~n}as-Garza, 2006]{branas2006poverty}
Bra{\~n}as-Garza, P. (2006).
\newblock Poverty in dictator games: {A}wakening solidarity.
\newblock {\em Journal of Economic Behavior \& Organization}, 60(3):306--320.

\bibitem[Bra{\~n}as-Garza, 2007]{branas2007promoting}
Bra{\~n}as-Garza, P. (2007).
\newblock Promoting helping behavior with framing in dictator games.
\newblock {\em Journal of Economic Psychology}, 28(4):477--486.

\bibitem[Bra{\~n}as-Garza et~al., 2016]{branasguess}
Bra{\~n}as-Garza, P., Rodriguez-Lara, I., and Sanchez, A. (2016).
\newblock Nobody expect selfishness.
\newblock {\em Mimeo}.

\bibitem[Brescoll, 2011]{brescoll2011who}
Brescoll, V.~L. (2011).
\newblock Who takes the floor and why: Gender, power, and volubility in
  organizations.
\newblock {\em Administrative Science Quarterly}, 56:622--641.

\bibitem[Camerer et~al., 2004]{camerer2004behavioural}
Camerer, C.~F., Ho, T.-H., and Chong, J.~K. (2004).
\newblock Behavioural game theory: Thinking, learning and teaching.
\newblock In {\em Advances in Understanding Strategic Behaviour}, pages
  120--180. Springer.

\bibitem[Capraro, 2015]{capraro2015emergence}
Capraro, V. (2015).
\newblock The emergence of hyper-altruistic behaviour in conflictual
  situations.
\newblock {\em Scientific Reports}, 4:9916.

\bibitem[Capraro and Marcelletti, 2014]{capraro2014good}
Capraro, V. and Marcelletti, A. (2014).
\newblock Do good actions inspire good actions in others?
\newblock {\em Scientific Reports}, 4:7470.

\bibitem[Charness and Gneezy, 2008]{charness2008s}
Charness, G. and Gneezy, U. (2008).
\newblock What's in a name? {A}nonymity and social distance in dictator and
  ultimatum games.
\newblock {\em Journal of Economic Behavior \& Organization}, 68(1):29--35.

\bibitem[Croson and Gneezy, 2009]{croson2009gender}
Croson, R. and Gneezy, U. (2009).
\newblock Gender differences in preferences.
\newblock {\em Journal of Economic Literature}, 47:448--474.

\bibitem[d'Adda et~al., 2015]{dadda2015push}
d'Adda, G., Capraro, V., and Tavoni, M. (2015).
\newblock Push, don't nudge: Behavioral spillovers and policy instruments.
\newblock {\em Mimeo}.

\bibitem[Dreber et~al., 2014]{dreber2014gender}
Dreber, A., von Essen, E., and Ranehill, E. (2014).
\newblock Gender and competition in adolescence: task matters.
\newblock {\em Experimental Economics}, 17(1):154--172.

\bibitem[Duflo, 2000]{duflo2000child}
Duflo, E. (2000).
\newblock Child health and household resources in south africa: Evidence from
  the old age pension program.
\newblock {\em The American Economic Review}, 90(2):393--398.

\bibitem[Dufwenberg and Muren, 2006]{dufwenberg2006gender}
Dufwenberg, M. and Muren, A. (2006).
\newblock Gender composition in teams.
\newblock {\em Journal of Economic Behavior \& Organization}, 61(1):50--54.

\bibitem[Eagly, 1987]{eagly1987sex}
Eagly, A.~H. (1987).
\newblock {\em Sex differences in social behavior: A social-role
  interpretation}.
\newblock Mahwah, NJ: Erlbaum.

\bibitem[Eckel and Grossman, 1998]{eckel1998women}
Eckel, C.~C. and Grossman, P.~J. (1998).
\newblock Are women less selfish than men?: {E}vidence from dictator
  experiments.
\newblock {\em The Economic Journal}, 108(448):726--735.

\bibitem[Eckel and Grossman, 2002]{eckel2002sex}
Eckel, C.~C. and Grossman, P.~J. (2002).
\newblock Sex differences and statistical stereotyping in attitudes toward
  financial risk.
\newblock {\em Evolution and Human Behavior}, 23(4):281--295.

\bibitem[Engel, 2011]{engel2011dictator}
Engel, C. (2011).
\newblock Dictator games: A meta study.
\newblock {\em Experimental Economics}, 14(4):583--610.

\bibitem[Fehr and Schmidt, 1999]{fehr1999theory}
Fehr, E. and Schmidt, K.~M. (1999).
\newblock A theory of fairness, competition, and cooperation.
\newblock {\em The Quarterly Journal of Economics}, 114:817--868.

\bibitem[Forsythe et~al., 1994]{forsythe1994fairness}
Forsythe, R., Horowitz, J.~L., Savin, N.~E., and Sefton, M. (1994).
\newblock Fairness in simple bargaining experiments.
\newblock {\em Games and Economic Behavior}, 6(3):347--369.

\bibitem[Heilman and Chen, 2005]{heilman2005same}
Heilman, M.~E. and Chen, J.~J. (2005).
\newblock Same behavior, different consequences: Reactions to men's and women's
  altruisitc citizenship behavior.
\newblock {\em Journal of Applied Psychology}, 90:431--434.

\bibitem[Heilman and Okimoto, 2007]{heilman2007why}
Heilman, M.~E. and Okimoto, T.~G. (2007).
\newblock Why are women penalized for success at male tasks?: The implied
  communality deficit.
\newblock {\em Journal of Applied Psychology}, 92:81--92.

\bibitem[Horton et~al., 2011]{horton2011online}
Horton, J.~J., Rand, D.~G., and Zeckhauser, R.~J. (2011).
\newblock The online laboratory: {C}onducting experiments in a real labor
  market.
\newblock {\em Experimental Economics}, 14(3):399--425.

\bibitem[Houser and Schunk, 2009]{houser2009fairness}
Houser, D. and Schunk, D. (2009).
\newblock Fairness, competition and gender: {E}vidence from {G}erman
  schoolchildren.
\newblock {\em Journal of Economic Psychology}, 30:634--641.

\bibitem[Kahneman et~al., 1986]{kahneman1986fairness}
Kahneman, D., Knetsch, J.~L., and Thaler, R.~H. (1986).
\newblock Fairness and the assumptions of economics.
\newblock {\em Journal of Business}, 59:S285--S300.

\bibitem[Lundberg et~al., 1997]{lundberg1997husbands}
Lundberg, S.~J., Pollak, R.~A., and Wales, T.~J. (1997).
\newblock Do husbands and wives pool their resources? {E}vidence from the
  united kingdom child benefit.
\newblock {\em Journal of Human Resources}, pages 463--480.

\bibitem[Mason and Suri, 2012]{mason2012conducting}
Mason, W. and Suri, S. (2012).
\newblock Conducting behavioral research on {A}mazon's {M}echanical {T}urk.
\newblock {\em Behavior Research Methods}, 44(1):1--23.

\bibitem[Niederle and Vesterlund, 2007]{niederle2007women}
Niederle, M. and Vesterlund, L. (2007).
\newblock Do women shy away from competition? {D}o men compete too much?
\newblock {\em The Quarterly Journal of Economics}, 122:1067--1101.

\bibitem[{Open Science Collaboration}, 2015]{open2015}
{Open Science Collaboration} (2015).
\newblock Estimating the reproducibility of psychological science.
\newblock {\em Science}, 349.

\bibitem[Paolacci and Chandler, 2014]{paolacci2014inside}
Paolacci, G. and Chandler, J. (2014).
\newblock Inside the {T}urk: {U}nderstanding {M}echanical {T}urk as a
  participant pool.
\newblock {\em Current Directions in Psychological Science}, 23(3):184--188.

\bibitem[Paolacci et~al., 2010]{paolacci2010running}
Paolacci, G., Chandler, J., and Ipeirotis, P.~G. (2010).
\newblock Running experiments on {A}mazon {M}echanical {T}urk.
\newblock {\em Judgment and Decision Making}, 5:411--419.

\bibitem[Rand et~al., 2016]{rand2016social}
Rand, D.~G., Brescoll, V., Everett, J.~A., Capraro, V., and Barcelo, H. (2016).
\newblock Social heuristics and social roles: {I}ntuition favors altruism for
  women but not for men.
\newblock {\em Journal of Experimental Psychology: General}.

\bibitem[Rubalcava et~al., 2009]{rubalcava2009investments}
Rubalcava, L., Teruel, G., and Thomas, D. (2009).
\newblock Investments, time preferences and public transfers paid to women.
\newblock {\em Economic Development and Cultural Change}, 57(3):507.

\end{thebibliography}
	

\pagebreak	
	
\section*{Figures}
	
\begin{figure}[H]
\caption{Expected behavior for men, women and both} \label{fig:exp}

\begin{subfigure}{0.5\textwidth}
\includegraphics[width=\linewidth]{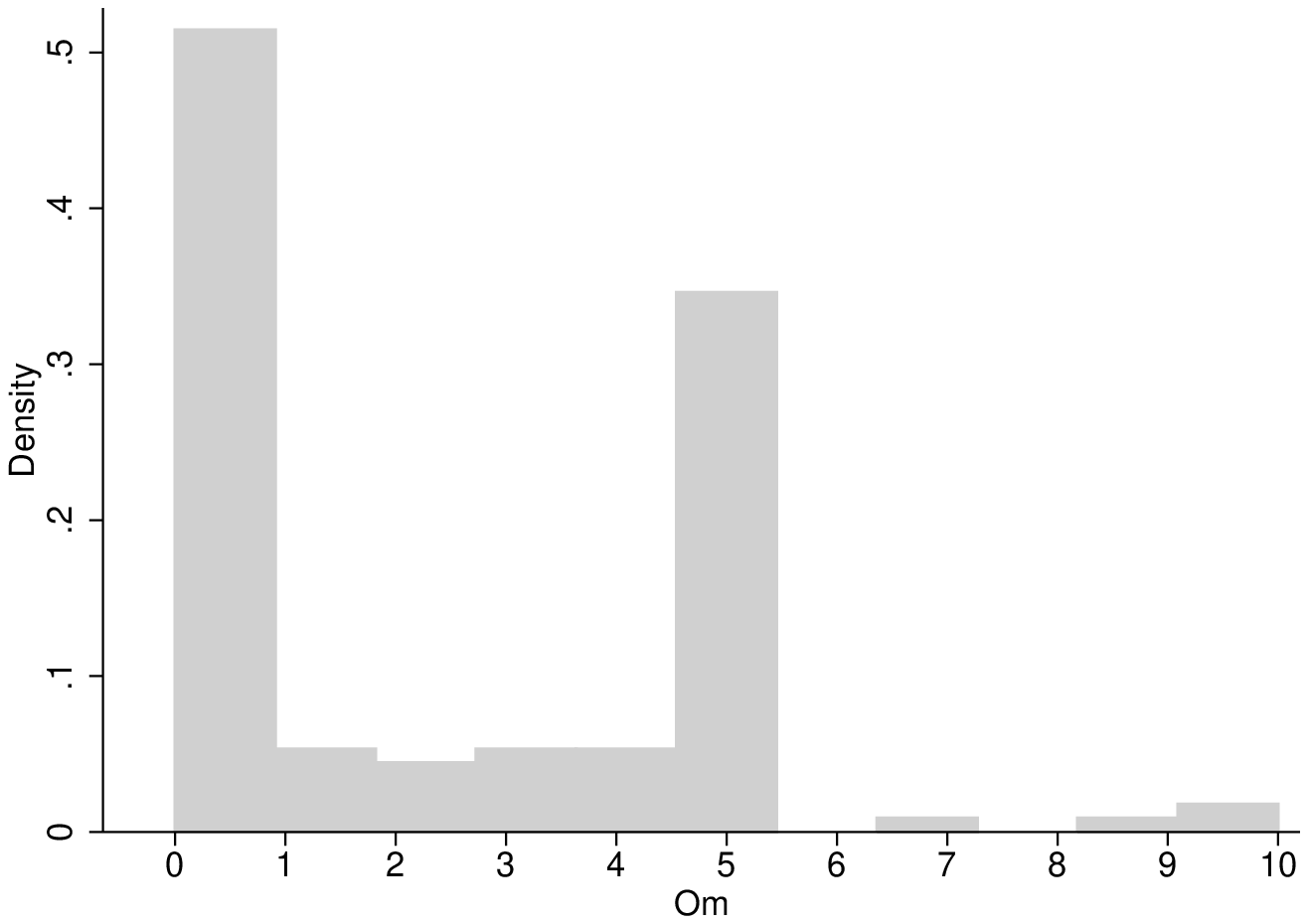}
\caption{Expectations for men [O$_{\text{m}}$]} \label{fig:1a}
\end{subfigure}
\hspace*{\fill} 
\begin{subfigure}{0.5\textwidth}
\includegraphics[width=\linewidth]{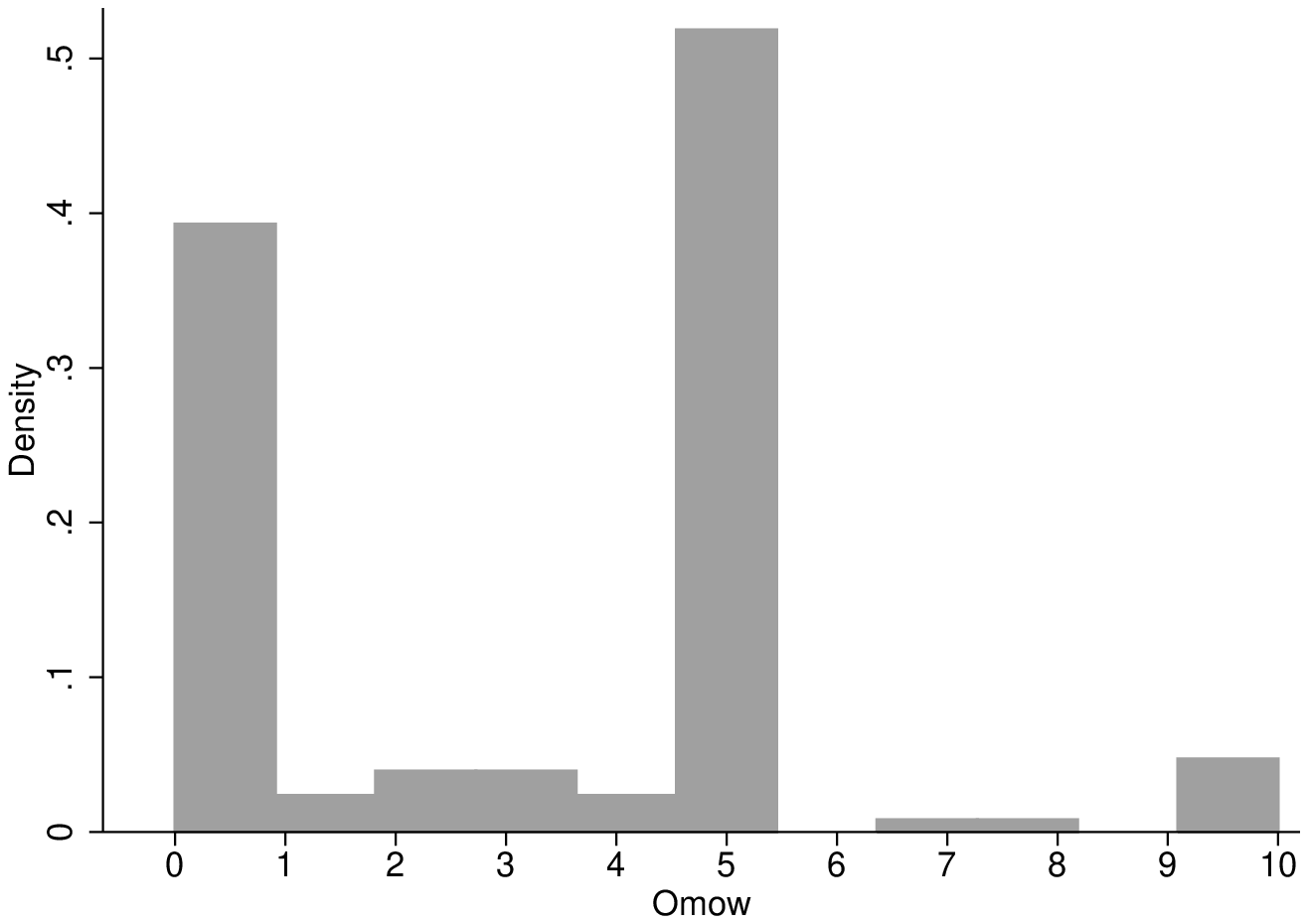}
\caption{Expectations for either men or women [O$_{\text{mow}}$]} \label{fig:1b}
\end{subfigure}\\
\begin{subfigure}{0.5\textwidth}
\includegraphics[width=\linewidth]{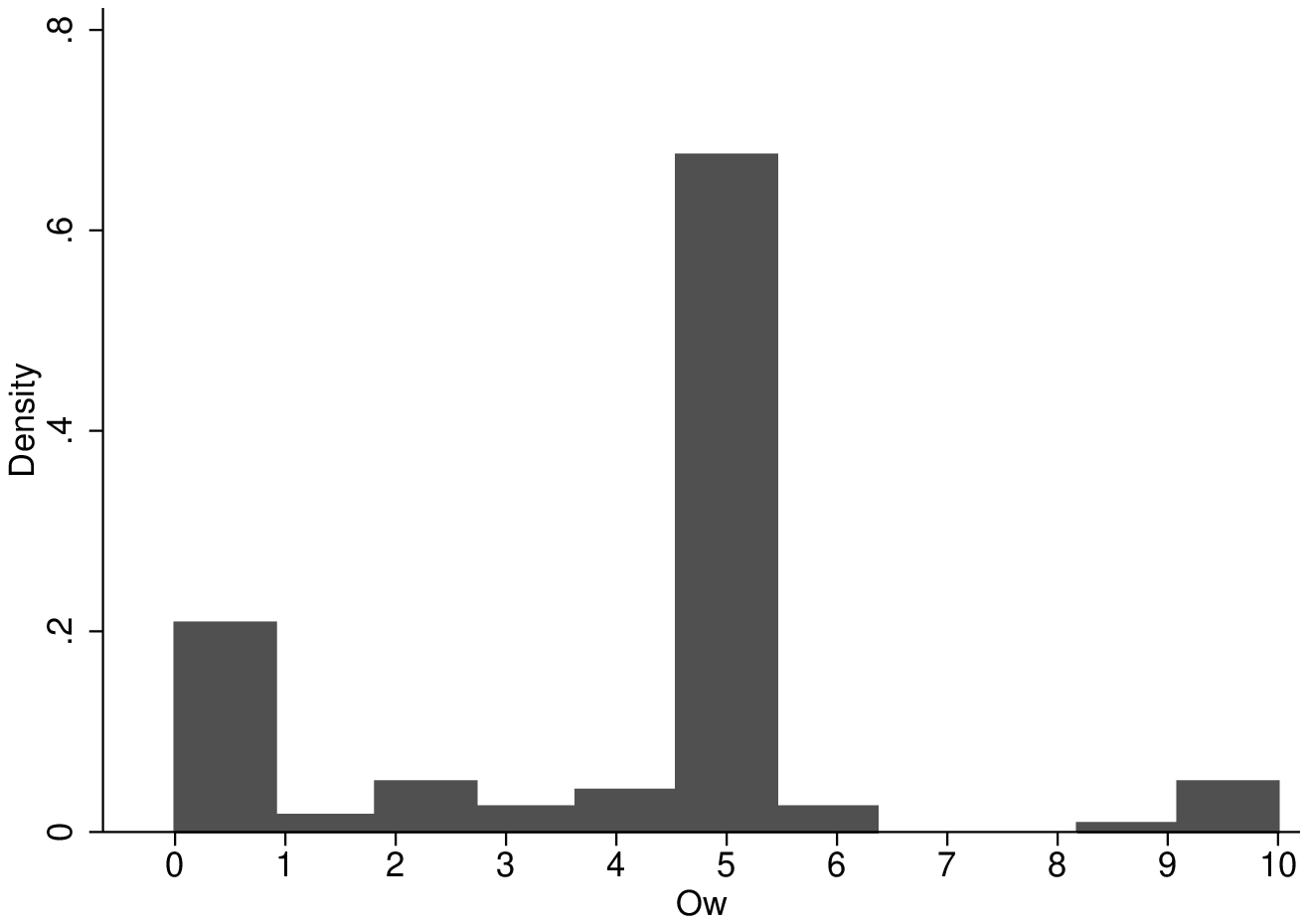}
\caption{Expectations for women [O$_{\text{w}}$]} \label{fig:1c}
\end{subfigure}
\hspace*{\fill} 
\begin{subfigure}{0.5\textwidth}
\includegraphics[width=\linewidth]{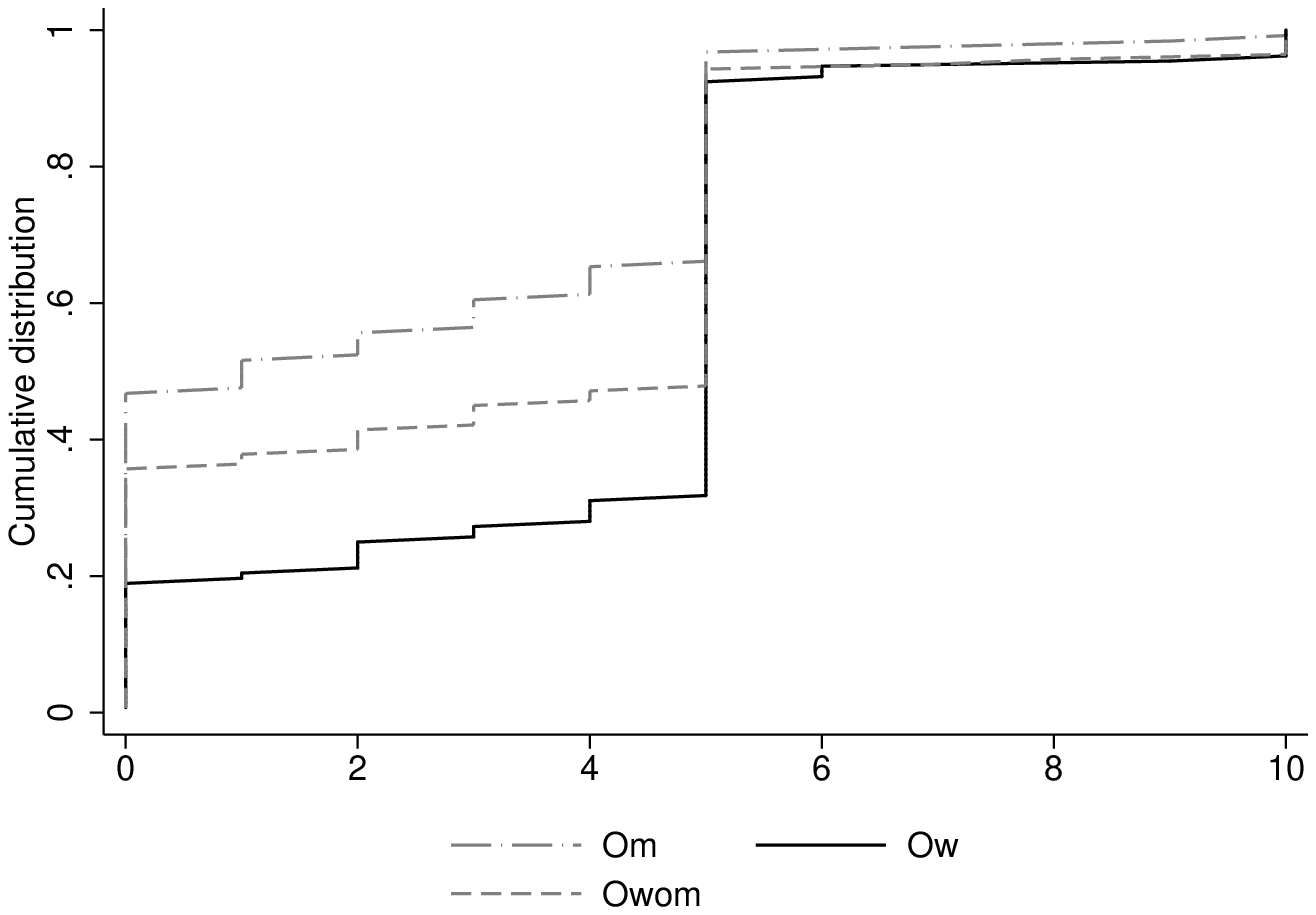}
\caption{CDFs for O$_{\text{m}}$, O$_{\text{mow}}$ and O$_{\text{w}}$} \label{T2vsT3.eps}
\end{subfigure}
\end{figure}
	
\pagebreak

\begin{figure}[H]
\caption{Actual behaviour: men vs women} \label{fig:2}

\begin{subfigure}{0.5\textwidth}
\includegraphics[width=\linewidth]{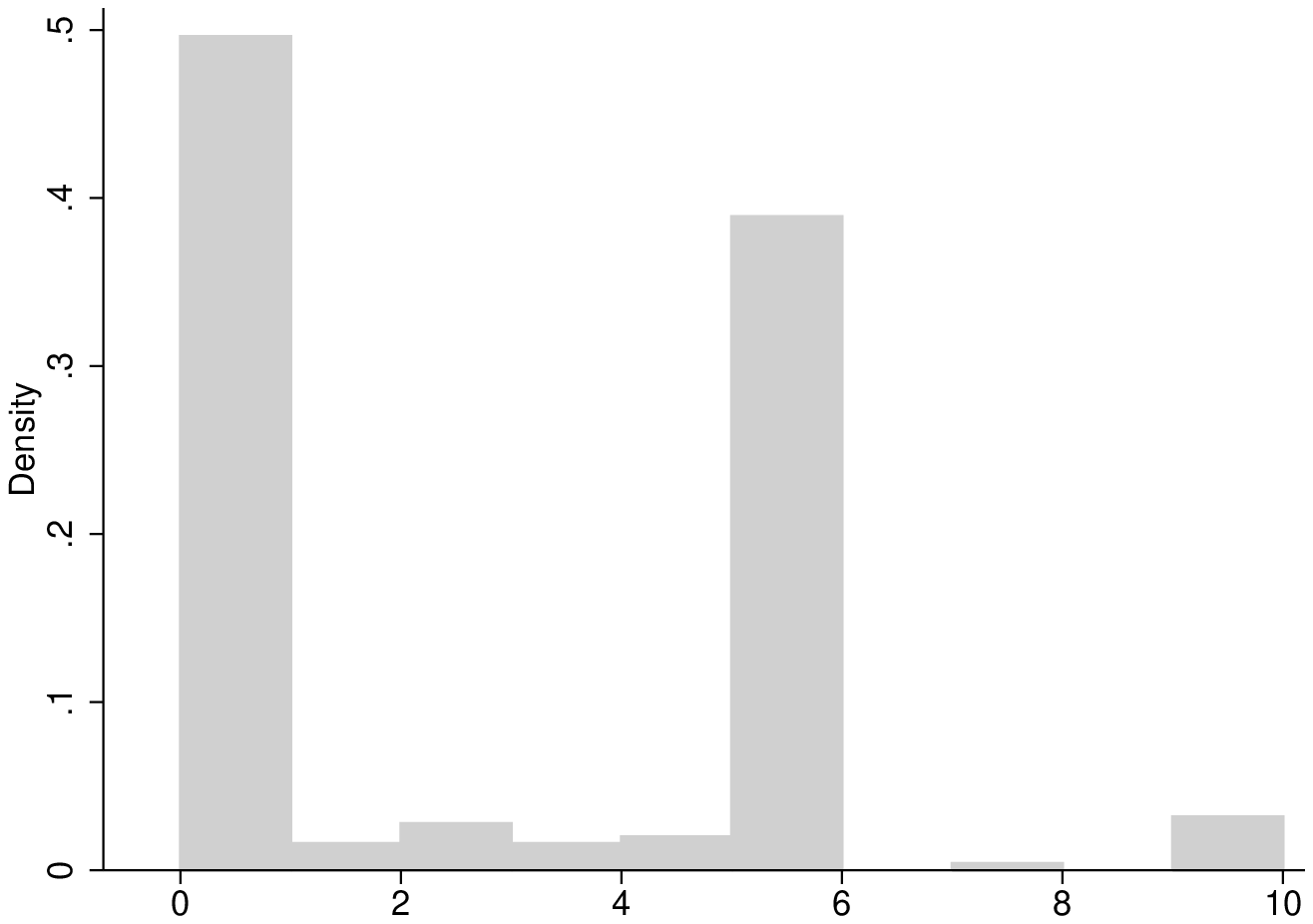}
\caption{Actual behaviour of men [Dm]} \label{fig:2a}
\end{subfigure}
\hspace*{\fill} 
\begin{subfigure}{0.5\textwidth}
\includegraphics[width=\linewidth]{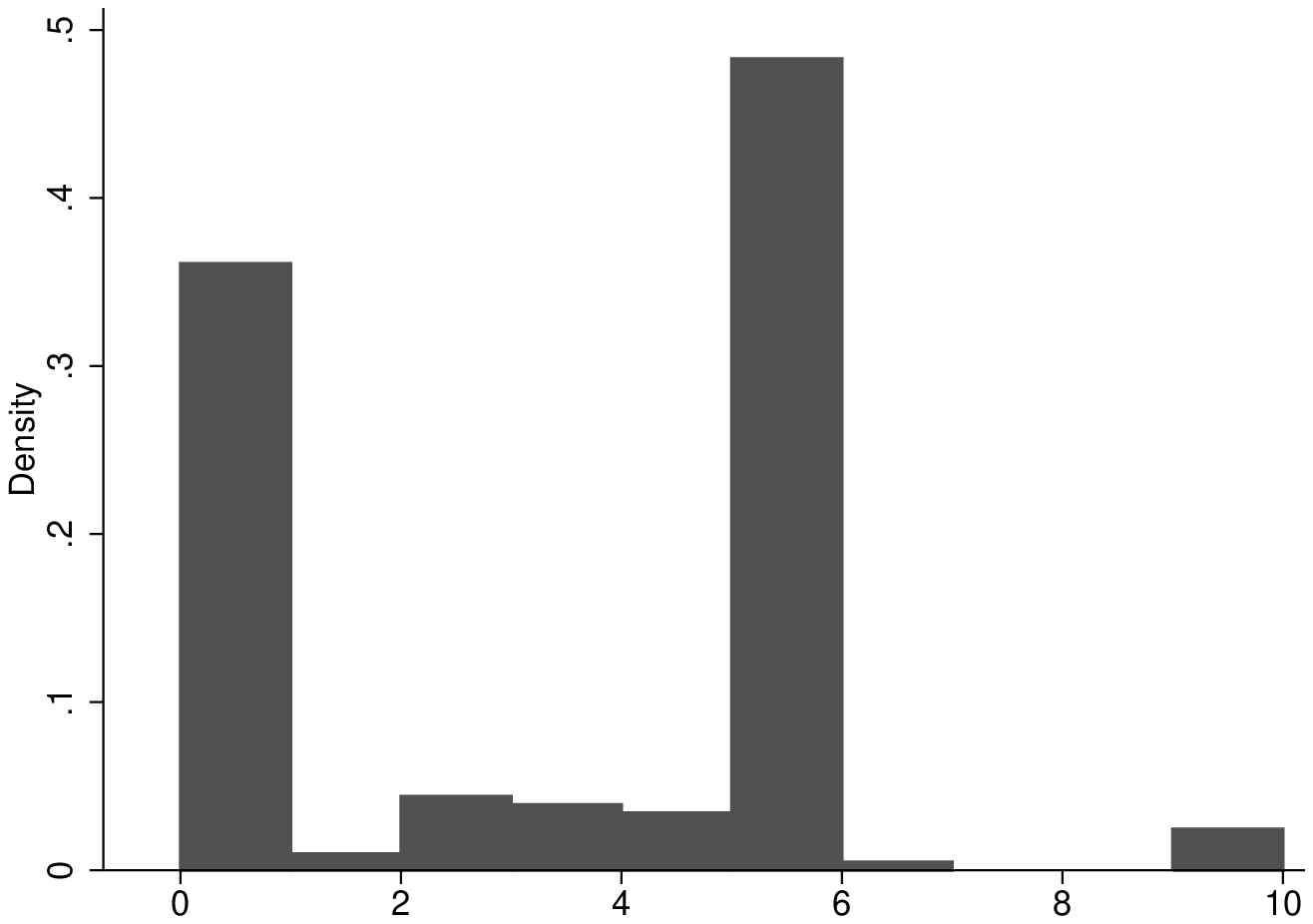}
\caption{Actual behaviour of women [Dw]} \label{fig:2b}
\end{subfigure}\\

\end{figure}

\pagebreak

\begin{figure}[H]
\caption{Accuracy of beliefs: Expected vs actual behaviour} \label{fig:2}

\begin{subfigure}{0.5\textwidth}
\includegraphics[width=\linewidth]{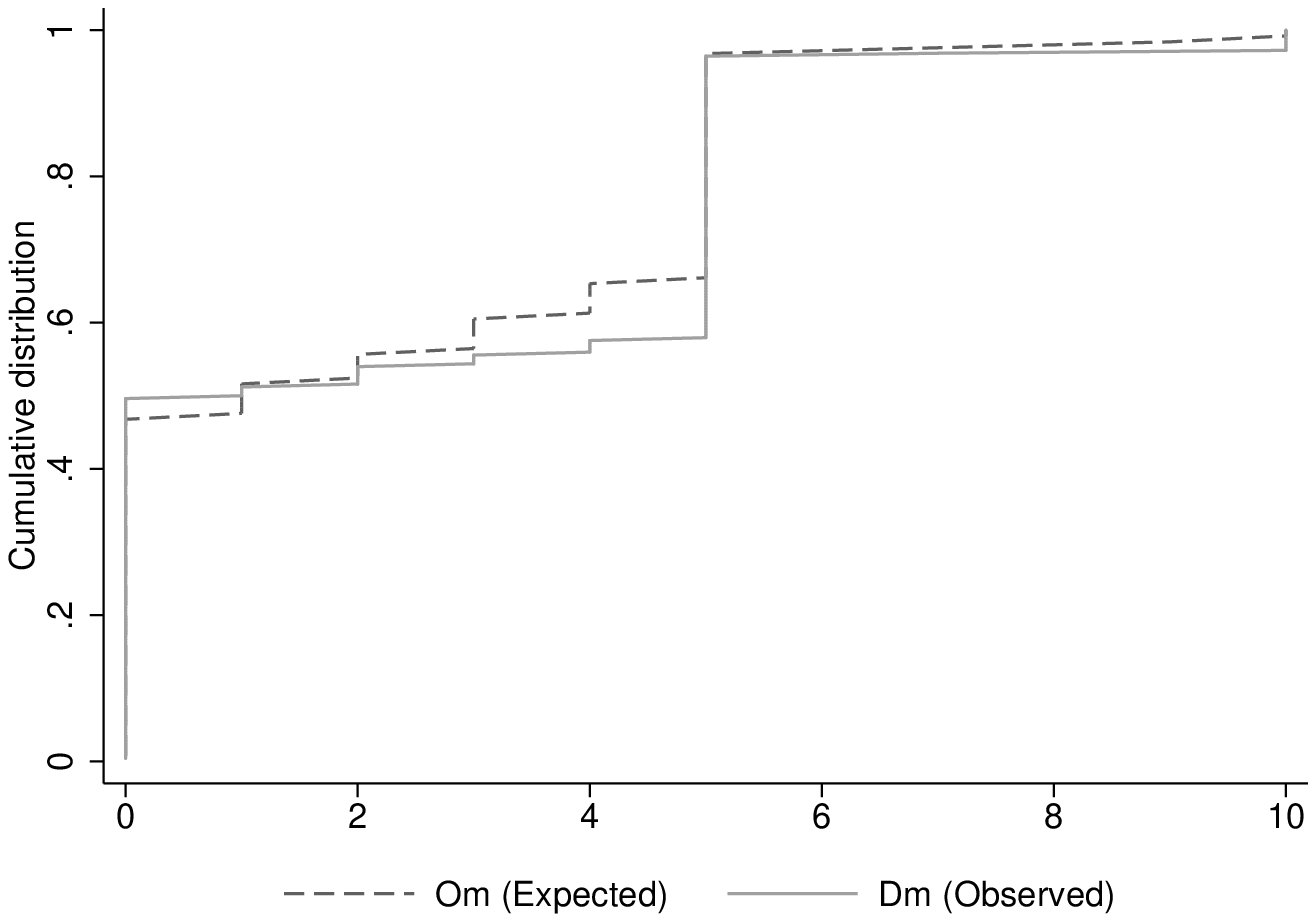}
\caption{Accuracy of beliefs for men} \label{fig:3a}
\end{subfigure}
\hspace*{\fill} 
\begin{subfigure}{0.5\textwidth}
\includegraphics[width=\linewidth]{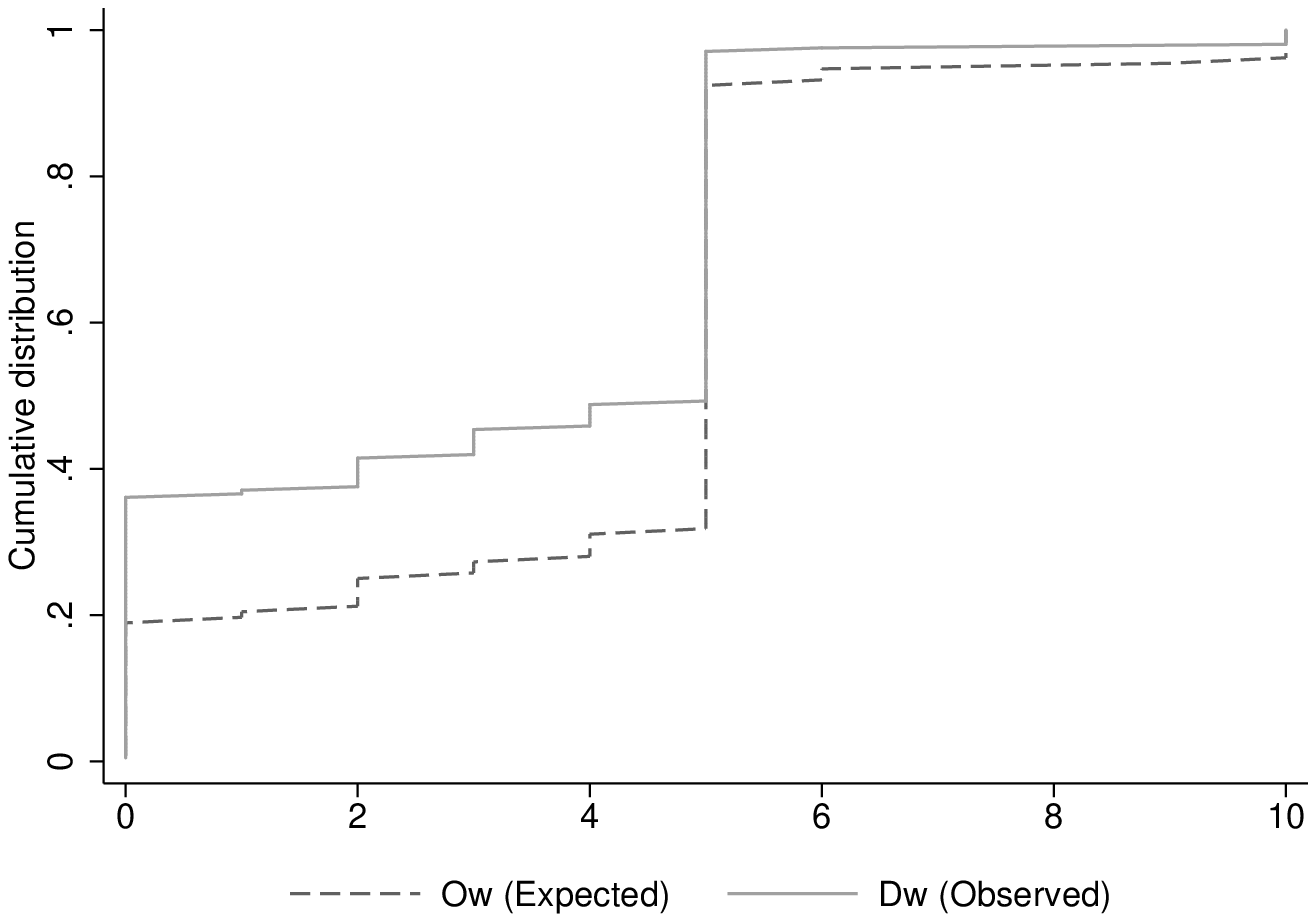}
\caption{Accuracy of beliefs for women} \label{T2vsT3.eps}
\end{subfigure}
\end{figure}

	\end{document}